\PassOptionsToPackage{dvipsnames}{xcolor}
\documentclass[10pt,conference]{IEEEtran}
\IEEEoverridecommandlockouts
\usepackage{cite}
\usepackage{amsmath,amssymb,amsfonts}
\usepackage{algorithmic}
\usepackage{graphicx}
\usepackage{textcomp}
\usepackage{xcolor}
\usepackage{balance}
\usepackage[all]{nowidow}

\usepackage[hidelinks]{hyperref}
\usepackage[capitalise]{cleveref}

\definecolor{darkblue}{rgb}{0.0, 0.0, 0.55}
\begin{document}

\title{Towards a Testbed for Scalable FaaS Platforms
    \thanks{Partially funded by the Bundesministerium f{\"u}r Bildung und Forschung (BMBF, German Federal Ministry of Education and Research) in the scope of the Software Campus 3.0 (Technische Universit\"at Berlin) program -- 01IS23068.}
}

\author{\IEEEauthorblockN{Trever Schirmer, David Bermbach}
    \IEEEauthorblockA{\textit{Technische Universit\"at Berlin}\\
        \textit{Scalable Software Systems Research Group} \\
        \{ts,db\}@3s.tu-berlin.de}
}

\maketitle

\begin{abstract}
    Most cloud platforms have a Function-as-a-Service (FaaS) offering that enables users to easily write highly scalable applications.
    To better understand how the platform's architecture impacts its performance, we present a research-focused testbed that can be adapted to quickly evaluate the impact of different architectures and technologies on the characteristics of scalability-focused FaaS platforms.
\end{abstract}


\section{Introduction}
\label{sec:introduction}
FaaS is a popular cloud computing paradigm that enables users to upload code that are executed in response to incoming events~\cite{Manner_2023_Definition}.
Today, popular cloud providers offers their own FaaS platform, each following different architectural approaches that enable highly scalable FaaS applications.
Additionally, open-source FaaS platforms such as OpenFaaS\footnote{\url{https://www.openfaas.com/}}, OpenWhisk\footnote{\url{https://openwhisk.apache.org/}} and Knative\footnote{\url{https://knative.dev/docs/}} exist.
These open-source platforms, however, often rely on using Kubernetes as underlying platform for managing function instances, which has been shown to limit scalability, leads to slow response times, and slow cold starts~\cite{Hima_2021_K8sScaling}.
Since the closed-source platforms from cloud providers are not easily adaptable for researchers, there is a gap in our understanding of highly scalable FaaS platforms: We can benchmark the existing closed-source platforms that are highly scalable with custom applications~\cite{schirmer2024fusionizepp}, but adapting the platform itself can only be done using open-source platforms.

To close this gap, we present our vision for a research-focused serverless platform with two goals: Its architecture should be as scalable as possible while making it easy to adapt to different use cases.
This enables evaluating the impact of specific changes to the platform under massive load.

Following the design of other open-source platforms, research-based FaaS platforms commonly use Kubernetes, resulting in the same problems other platforms face.
Other research platforms are designed for lower load, which does not necessarily translate to highly scalable platforms~\cite{Liu_2023_Gap}.
Additionally, researchers often write new platforms to test out ideas instead of adapting existing ones~\cite{Wen_2023_RiseOfThe}, likely due to the limitations of available open-source platforms.
This, however, increases the difficulty of comparing them.
In this paper, we first present the architecture of a highly scalable testbed that can be used for multiple research use cases and then present different research questions that can be answered using it.

\begin{figure}
    \centering
    \includegraphics[width=\linewidth]{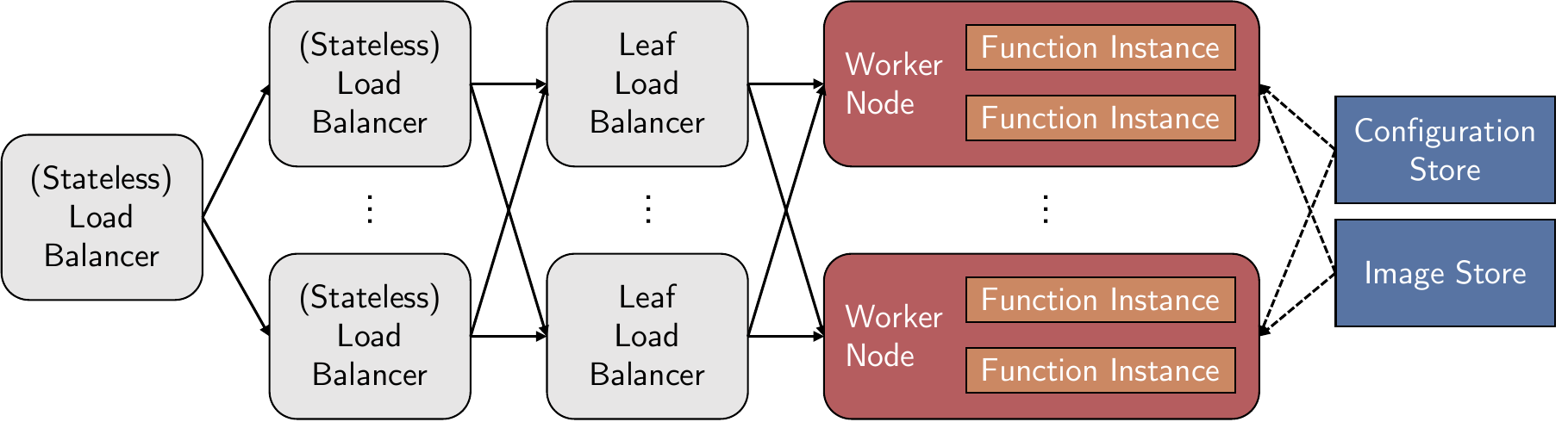}
    \caption{\textit{HyperFaaS} Architecture.
    Requests enter through a series of stateless or stateful load balancers that can use different algorithms to assign requests to worker nodes.
    Worker nodes start and stop function instances on-demand and can access configuration and image stores to get information about unknown functions.
    The architecture is designed to be adaptable: Different types of load balancers, function instances, and billing models can easily be implemented.
    }
    \label{fig:overview}
\end{figure}

\section{HyperFaaS}

The testbed, called HyperFaaS, is currently under development.
The overall architecture is shown in \cref{fig:overview} and follows a tree-like pattern.
The root of the tree consists of load balancers, which can follow different algorithms, distributing requests through the tree.
They all have the same interface for calling functions.
This is the core of what makes the system scalable: To scale the system up by a factor of two, simply replicate the existing servers and add load balancer in front to randomly assign requests to one branch.

The leaf nodes have the same interface as the other load balancers, but instead of calling other load balancers, they call worker nodes.
Leaf nodes can call multiple worker nodes, and worker nodes can be called by different leaf nodes in parallel.
While the leaf nodes can be stateless, they can access the current state of the workers to efficiently route requests (e.g., to workers that already have an instance of the function).

The worker nodes start function instances based on incoming requests and stop them after a timeout or after they are stopped by a leaf node.
Functions can be configured to allow a certain amount of parallel requests in the same function instance, and worker nodes keep track of the amount of in-flight requests for a function instance.

Finally, an image registry is used to distribute function images to worker nodes, and a configuration store keeps track of the resource limits of functions.
These need to be read by the worker nodes to start instances and can also be read by load balancers using more complicated scheduling algorithms.
As the focus of this testbed is the FaaS platform itself, we assume that the cloud platform itself already offers object storage for the images and a key-value store for the configuration that can scale with the demands of the platform.

\subsubsection*{Prototype}

The testbed, which is still under active development, is written in Go and available as open-source.\!\footnote{\url{https://github.com/3s-rg-codes/HyperFaaS}}
Communication is done via gRPC, a popular RPC middleware with support for many programming languages, enabling components to be easily switched out.
While the method of running function instances is also replaceable, the current implementation uses Docker containers.

\section{Research Questions}

The testbed can be used to quickly write prototypes for research questions.
As examples of the kinds of questions that can be answered, we present two open research questions we are currently working on.

\subsection{Within-Instance Concurrency}

Different FaaS platforms follow different approaches for how many requests can be routed to the same function instance at the same time (i.e, the concurrency of requests within one instance).
For example, AWS Lambda\footnote{\url{https://aws.amazon.com/lambda/}} only allows a within-instance concurrency of just one.
Other platforms such as Azure Functions\footnote{\url{https://azure.microsoft.com/en-us/products/functions}}, Google Cloud Run Functions\footnote{\url{https://cloud.google.com/functions}}, and many Kubernetes-based implementations instead have unlimited within-instance concurrency.
Instead, they periodically inspect the resource consumption of the instance and add additional replicas if this limit is exceeded.
The Kubernetes-based FaaS platform Knative has support for setting any hard limit on the within-instance concurrency.

Today, to research the impact of within-instance concurrency on the overall platform behavior (e.g., the additional scaling potential) requires comparing entirely different platforms, which has many disruptive influences.
With this testbed, it is possible to simply switch out the load balancers to keep track of within-instance concurrency (or not), enabling a fair comparison between different approaches.

\subsection{Emulating Worker Nodes for Benchmarking}

\begin{figure}
    \centering
    \includegraphics[width=0.7\linewidth]{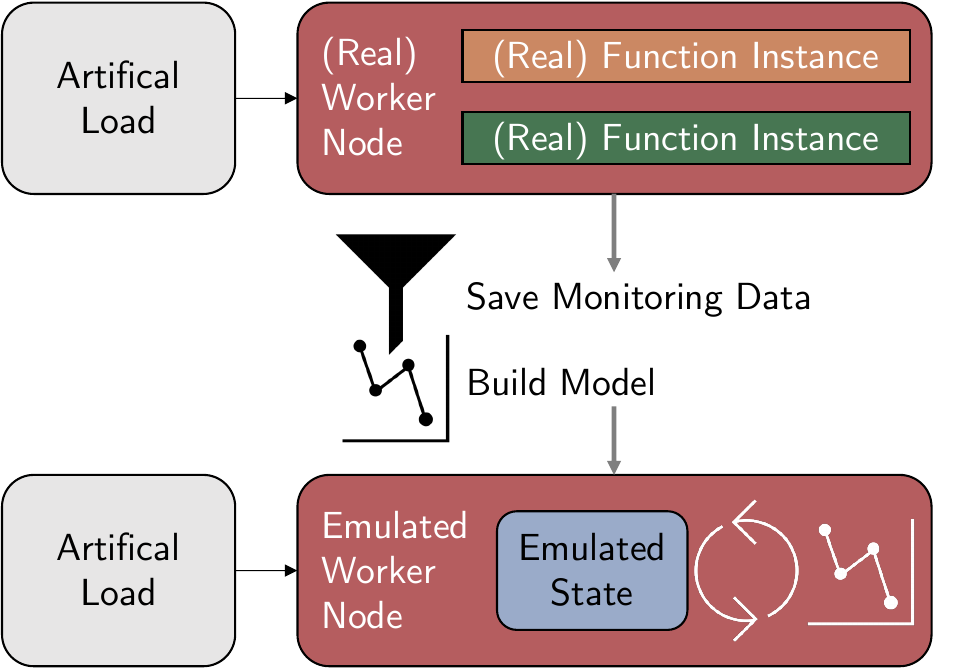}
    \caption{
        To decrease the resource consumption of the workers for benchmarks where their actual execution is not needed, it is possible to emulate them instead.
        As a first step, an actual server is put under artificial load.
        The resulting monitoring data is saved and used to build a model of the worker node, e.g., using machine learning.
        To evaluate the emulated worker performance, the same load that was used in step one can be used to measure how similarly it behaves.
        Doing this enables running many more emulated workers using the same resources.
    }
    \label{fig:hyperfake}
\end{figure}

Benchmarking a FaaS platform requires calling many functions with high concurrency.
The main computational load on the platforms then stems from the workers starting function instances and actually executing the requests.
There are, however, some research questions where the actual result of the computation is not important.
For example, when trying out different load balancing strategies (e.g., stateless vs. stateful), the result of the computation is not needed.
Instead, it is enough to have a component that behaves as if it were a worker node to the outside, without actually executing user code.
Whenever a function is called on this emulated worker, it should have the same kind of answer within the same timeframes with a comparable failure rate.
To achieve this, we plan to use the pipeline shown in~\cref{fig:hyperfake}: First, we run one server as a worker node and put it under artificial load with all functions that will be used in the final experiment.
The metrics of all invocations and the server utilization are saved, and a model of the worker node is built based on this data.
This model can either be a simple linear regression model, or a more complicated model using machine learning.
Based on the model, the same server that was used as worker node in step one can then be re-used to run many emulated servers.
Overall, this allows researchers to allocate more resources to the components that are actually being evaluated, while fewer resources need to be used to emulate many worker nodes.

\section{Conclusion}

In this paper, we presented our vision of a testbed for scalable FaaS platforms called HyperFaaS.
We then presented use cases that we are currently working on, which will be enabled by this testbed.

\bibliographystyle{IEEEtran}
\bibliography{bibliography}

\begin{thebibliography}{1}
\providecommand{\url}[1]{#1}
\csname url@samestyle\endcsname
\providecommand{\newblock}{\relax}
\providecommand{\bibinfo}[2]{#2}
\providecommand{\BIBentrySTDinterwordspacing}{\spaceskip=0pt\relax}
\providecommand{\BIBentryALTinterwordstretchfactor}{4}
\providecommand{\BIBentryALTinterwordspacing}{\spaceskip=\fontdimen2\font plus
\BIBentryALTinterwordstretchfactor\fontdimen3\font minus \fontdimen4\font\relax}
\providecommand{\BIBforeignlanguage}[2]{{%
\expandafter\ifx\csname l@#1\endcsname\relax
\typeout{** WARNING: IEEEtran.bst: No hyphenation pattern has been}%
\typeout{** loaded for the language `#1'. Using the pattern for}%
\typeout{** the default language instead.}%
\else
\language=\csname l@#1\endcsname
\fi
#2}}
\providecommand{\BIBdecl}{\relax}
\BIBdecl

\bibitem{Manner_2023_Definition}
\BIBentryALTinterwordspacing
J.~Manner, ``A structured literature review approach to define serverless computing and function as a service,'' in \emph{2023 IEEE 16th International Conference on Cloud Computing (CLOUD)}, Jul. 2023, p. 516–522. [Online]. Available: \url{https://ieeexplore.ieee.org/document/10255020/}
\BIBentrySTDinterwordspacing

\bibitem{Hima_2021_K8sScaling}
H.~Govind and H.~González–Vélez, ``Benchmarking serverless workloads on kubernetes,'' in \emph{2021 IEEE/ACM 21st International Symposium on Cluster, Cloud and Internet Computing (CCGrid)}, 2021, pp. 704--712.

\bibitem{schirmer2024fusionizepp}
\BIBentryALTinterwordspacing
T.~Schirmer, J.~Scheuner, T.~Pfandzelter, and D.~Bermbach, ``Fusionize++: Improving serverless application performance using dynamic task inlining and infrastructure optimization,'' \emph{IEEE Transactions on Cloud Computing}, Aug. 2024. [Online]. Available: \url{https://doi.org/10.1109/TCC.2024.3451108}
\BIBentrySTDinterwordspacing

\bibitem{Liu_2023_Gap}
\BIBentryALTinterwordspacing
Q.~Liu, D.~Du, Y.~Xia, P.~Zhang, and H.~Chen, ``The gap between serverless research and real-world systems,'' in \emph{Proceedings of the 2023 ACM Symposium on Cloud Computing}, ser. SoCC ’23.\hskip 1em plus 0.5em minus 0.4em\relax New York, NY, USA: Association for Computing Machinery, Oct. 2023, p. 475–485. [Online]. Available: \url{https://dl.acm.org/doi/10.1145/3620678.3624785}
\BIBentrySTDinterwordspacing

\bibitem{Wen_2023_RiseOfThe}
J.~Wen, Z.~Chen, X.~Jin, and X.~Liu, ``\BIBforeignlanguage{en}{Rise of the planet of serverless computing: A systematic review},'' \emph{\BIBforeignlanguage{en}{ACM Transactions on Software Engineering and Methodology}}, p. 3579643, Jan. 2023.

\end{thebibliography}

\end{document}